\documentclass[iop]{emulateapj}
\usepackage{epsfig}
\usepackage{csquotes}
\usepackage{graphicx}
\usepackage{ulem,color}
\usepackage[dvipsnames]{xcolor/xcolor}
\usepackage{dcolumn}
\usepackage{bm}
\usepackage{longtable}
\usepackage{natbib}
\usepackage[bookmarks=true]{hyperref}
\usepackage{xspace}
\usepackage{amsmath}
\newcommand{\harm}{{\sc Harm3d}\xspace}

\def\lesssim{\mathrel{\hbox{\rlap{\hbox{\lower4pt\hbox{$\sim$}}}\hbox{$<$}}}}
\def\gtrsim{\mathrel{\hbox{\rlap{\hbox{\lower4pt\hbox{$\sim$}}}\hbox{$>$}}}}

\def\lambdabar{%
\relax
\bgroup
\def\@tempa{\hbox{\raise.73\ht0
\hbox to0pt{\kern.25\wd0\vrule width.5\wd0
height.1pt depth.1pt\hss}\box0}}%
\mathchoice{\setbox0\hbox{$\displaystyle\lambda$}\@tempa}%
{\setbox0\hbox{$\textstyle\lambda$}\@tempa}%
{\setbox0\hbox{$\scriptstyle\lambda$}\@tempa}%
{\setbox0\hbox{$\scriptscriptstyle\lambda$}\@tempa}%
\egroup
}

\slugcomment{Accepted for publication in ApJL}

\begin{document}

\title{Quasi-Periodic Behavior of Mini-Disks in Binary Black Holes Approaching Merger}

\author{Dennis B. Bowen    $^1$,
        Vassilios Mewes    $^1$,
        Manuela Campanelli $^1$,\\
        Scott C. Noble        $^{2,3}$,
        Julian H. Krolik      $^4$,                
        and Miguel Zilh\~ao   $^5$
} 

\affil{$^1$ Center for Computational Relativity and Gravitation, 
  Rochester Institute of Technology, Rochester, NY 14623\\
  $^2$ Department of Physics and Engineering Physics,
  The University of Tulsa, Tulsa, OK 74104\\
  $^3$ NASA Postdoctoral Program Senior Fellow, Goddard Space Flight Center, Greenbelt, MD 20771\\
  $^4$ Department of Physics and Astronomy, Johns Hopkins
  University, Baltimore, MD 21218\\
  $^5$CENTRA, Departamento de F\'isica, Instituto Superior T\'ecnico, Universidade de Lisboa,
  1049 Lisboa, Portugal}

\email{dbbsma@rit.edu}

\begin{abstract}
We present the first magnetohydrodynamic simulation in which a circumbinary
disk around a relativistic binary black hole feeds mass to individual accretion
disks (\enquote{mini-disks}) around each black hole. 
Mass flow through the accretion streams linking the circumbinary disk to
the mini-disks is modulated quasi-periodically by the streams' interaction with a nonlinear $m=1$ density
feature, or \enquote{lump}, at the inner edge of the circumbinary disk:
the stream supplying each mini-disk comes into phase with the lump
at a frequency $0.74$ times the binary orbital frequency.
Because the binary is relativistic, the tidal truncation radii of
the mini-disks are not much larger than their innermost stable circular orbits;
consequently, the mini-disks' inflow times are shorter than the conventional estimate
and are comparable to the stream modulation period.
As a result, the mini-disks are always in inflow disequilibrium, with their
masses and spiral density wave structures responding to
the stream's quasi-periodic modulation.
The fluctuations in each mini-disk's mass are so large that as much as
$75\%$ of the total mini-disk mass can be contained within a single mini-disk. 
Such quasi-periodic modulation of the
mini-disk structure may introduce distinctive time-dependent
features in the binary's electromagnetic emission.
\end{abstract}


\section{Introduction}
\label{sec:introduction}

Supermassive binary black holes (SMBBHs) are expected to form
following galaxy mergers (see~\cite{Khan16} and~\cite{Kelley17} for
recent work).  
When copious gas is available, SMBBHs may be
luminous electromagnetically as a result of accretion.
Newtonian simulations including the black holes (BHs) in the
computational domain have demonstrated that the accreting material
from the circumbinary disk forms individual mini-disks around each
BH~\citep{RoedigSesana2012,Farris14,Farris15,Farris15a,MunozLai16,Tang17,Tang18}.
The spectra of these systems may exhibit distinctive features,
both in the infrared/optical range and in hard X-rays~\citep{Roedig:2014}.
It is also important to predict the character of their electromagnetic emission
later in the binary's history, as it approaches merger, both for its intrinsic
interest and because the behavior of gas in this period creates the conditions
upon which the merger acts.  Unfortunately, simulations of matter flow from a
circumbinary disk to a BBH system in the relativistic regime have
been unable to form any such 
mini-disks~\citep{2010ApJ...715.1117B,Pal10,Farris11,Bode12, Farris12,Giacomazzo12,Gold14}.
However, if mini-disks are placed in such a system, hydrodynamic simulations
have found that they remain stable down to a binary separation $a{\lesssim}20M$\footnote{
We adopt geometrized units with $G=c=1$ 
where $r_g \equiv GM/c^2$ and $M$ is the total mass of the binary.}~\citep{Bowen17}.

In this paper, we present the first-ever
general relativistic magnetohydrodynamic (MHD) simulation in which matter is accreted
from a circumbinary disk into the domain of a relativistic BBH with mini-disks
around the BHs.  We observe a dynamic coupling between
the accretion streams linking the circumbinary disk to the mini-disks
and a $m=1$ modulation of the gas density at the inner edge of the circumbinary disk,
a \enquote{lump}.  Each BH comes into phase with the lump at a rate equal to the  beat
frequency between the lump's orbital frequency and the binary orbital frequency $(\Omega_{\rm{bin}})$,
$\approx 0.74\Omega_{\rm{bin}}$; while in phase, a BH's accretion stream
carries more material than the other BH's stream.  The stronger stream
then alternates from one BH to the other.

This lump-driven asymmetric accretion has been observed in
previous simulations
\citep{Shi12,Noble12,DOrazio13,Farris14,Farris15,Farris15a,DOrazio16},
but when the binary separation is relativistic it leads
to more dramatic consequences. Newtonian studies with two
mini-disks~\citep{Farris14,Farris15,Farris15a} produced
nearly identical mini-disks for equal-mass binaries.
By contrast, in these relativistic binaries the relatively modest
distance from a mini-disk's tidal truncation radius to its innermost stable circular
orbit (ISCO) implies an inflow time comparable to the stream alternation period;
as a result, the mini-disks' internal structures, accretion rates and masses vary strongly
on the alternation timescale, keeping the mini-disks in a permanent state of
inflow disequilibrium.
\clearpage

\section{Simulation Details}
\label{sec:simulation-details}
\subsection{General Relativistic Magnetohydrodynamics}
The aim of our simulation is to investigate the dynamics of
individual mini-disks due to their interaction with the accretion from
the inner edge of the circumbinary disk when
the binary's orbit is relativistic.  We approximate
the spacetime of the binary by asymptotically
matching BH perturbation theory to post-Newtonian (PN) theory;
this description satisfies the Einstein Field Equations well in the  
regime of this study (for full details see~\cite{Noble12,PROJ0,Ireland:2015cjj,Bowen17}). 
We evolve the equations of general relativistic MHD on this background spacetime in flux-conservative
form using the \harm code (see~\cite{Noble09,Noble12,Bowen17}), neglecting 
the self-gravity of the gas because the gas mass is negligible
in comparison to the SMBBH's mass for astrophysically relevant systems.

The gas's thermodynamics is governed by an adiabatic 
equation of state and local cooling. We cool towards
the system's initial local entropy via the prescription of~\citet{Noble12}. 
Following~\cite{Bowen17}, the cooling time is set differently in each of four
distinct regions; one for the circumbinary region $(r>1.5a)$, one
for each mini-disk $(r_i<0.45a)$, and one for the cavity between the
mini-disk and circumbinary regions. Here $r$ and $r_i$ denote the
distance to the center-of-mass and individual $i^{th}$ BH,
respectively. The cooling time is taken to be the local, equatorial
Keplerian orbital period of the fluid in the circumbinary and
mini-disk regions. Everywhere else, the cooling time is taken as that
for $r = 1.5a$. For full details on the calculations of these cooling
times see~\cite{Bowen17}.

\subsection{Grid and Boundary Conditions}

We perform our simulation in a dynamic, double fish-eye (warped)
spherical PN harmonic coordinate system whose
origin lies at the center-of-mass~\citep{WARPED}.  This system
concentrates cells in the immediate vicinity of the BHs in order to resolve
the mini-disks, while smoothly transitioning to spherical coordinates in the
circumbinary disk.  See~\cite{WARPED,Bowen17} for further insight into the grid 
structure employed.  

We excise a sphere of radius $r=2M$ around the origin at the center-of-mass
because the warped coordinates are topologically spherical.
Although this cut-out removes material sloshing between the two mini-disks~\citep{Bowen17},
an excision region this large enlarges the time-step sufficiently for this
simulation to run within our available resources.
We impose outflow boundary conditions on the ``radial" $x^1$ boundaries,
requiring that the radial component of the 4-velocity $u^r$
be either zero or oriented out of the domain on the boundary.  The outer
boundary is located at $13a_0$, where $a_0$ is the initial binary
separation. We apply reflective axisymmetric 
boundary conditions at the ``polar angle" $x^2$
boundaries and periodic boundary conditions at the ``azimuthal"
$x^3$ boundaries.

We use $600{\times}160{\times}640$ ($x^1{\times}x^2{\times}x^3$) cells. The equatorial
plane cell counts were chosen to follow~\citet{Bowen17}. 
The poloidal cell count does not fully meet the standard of 32 cells per scale height
in the outer portions of the mini-disks on the side farthest from the center-of-mass,
but it allows the simulation to be accessible with our available resources;
in the circumbinary disk, we use a poloidal grid identical to that of~\cite{Noble12},
who demonstrated that it provides good quality resolution.

\subsection{Initial Data}
We initialize our simulation using two separate regions, one
for the domain of the mini-disks $(r<a_0)$ and another for
the circumbinary data. The initial state of the circumbinary disk
is the equilibrated state of RunSE in~\cite{Noble12} at $t=50,000M$.
This time was selected to ensure that the disk had sufficiently equilibrated
in terms of MHD turbulence and azimuthally-averaged
surface density, and had also begun to break axisymmetry near its
inner edge. 
The circumbinary data were obtained from a simulation that employed
a different coordinate basis and grid discretization from the one
used for our present simulation, so we needed to perform a tri-linear
interpolation onto our grid. However, this interpolation step
introduces violations to the magnetic field divergence constraint.

We remove the divergence from, or \enquote{clean},
our solution by employing a projection method~\citep{brackbill1980effect,2000JCoPh.161..605T}
 until $\approx99\%$ of the domain reaches a target of
\begin{equation}
  \nabla\cdot\mathbf{B}\leq10^{-3}b_\lambda b^\lambda / dx^i_{\mathrm{min}},
\end{equation}
where $b_\lambda$ is the magnetic 4-vector and
$dx^i_{\mathrm{min}}$ is the smallest cell dimension anywhere in the grid.

For the mini-disk region $(r<a_0)$, we initialize nearly hydrostationary torii
around each BH, neglecting the presence of the binary companion~\citep{Bowen17}. 
For this study, we extended the method to include an initial seed magnetic field
and the vertical structure of the disks. In the local Boyer-Lindquist frame of each
BH we set the time, radial, and polar angle components of the vector potential to
zero, but set 
\begin{equation}
  A_{\phi_{BL}}=\mathrm{max}\left(\bar{\rho}-\frac{1}{4}\rho_{\rm{max}},\,0\right),
\end{equation}
where $\bar{\rho}$ is the average density of a cell and its 
neighbors and $\rho_{\rm{max}}$ is the maximum density of the mini-disk.
$A_\mu$ is then transformed into and interpolated onto the warped grid.
Finally, we calculate the magnetic field and normalize it to produce the desired initial
plasma $\beta$. The full details of the mini-disk initialization is described in~\cite{Bowen17}.

We select the mini-disk initial parameters to be $r_{\rm{in}}=3.1M$, $r_{\rm{pmax}}=4.6M$,
$\beta=100$, $H/r_{\rm{pmax}}=0.09$
($H$ is the vertical scale height), and initial entropy
$S_{0}{\equiv}P/\rho^{5/3}=0.01$.  These values were chosen so that
in their initial state the mini-disks were wholly outside the ISCO and had
structural properties, but not necessarily inflow rates, similar to
those of the initial circumbinary disk.


\section{Results}
\label{sec:results}
Our simulation starts with an initial binary separation $a_0=20M$ 
(defined in the PN harmonic gauge),
a distance chosen so that low-order (1PN) effects
are $\sim O(0.1)$ and binary inspiral due to gravitational radiation is noticeable,
but small, over the duration of a simulation.
As observed in~\cite{Bowen17}, there is a brief initial transient in which
the mini-disks expand because tidal forces were neglected in their construction.
After about a binary orbital period, the mini-disks become tidally truncated
at $r_t\simeq0.3a$ and begin to accrete onto their individual BHs.  Meanwhile, the
circumbinary disk delivers mass to the mini-disks through a pair of accretion streams
that forms almost immediately.

As can be readily gleaned from Figure~\ref{fig:snapshots}, the two
accretion streams are not symmetric, and the contrast in their mass content grows
with time.  The more massive stream always has its origin in the lump at the inner
edge of the circumbinary disk.  Due to this asymmetry, the streams preferentially
deposit material onto one of the mini-disks at any given time, despite the BH's
equal masses.  We examine this effect in greater detail in the next subsection.

\subsection{Asymmetric Accretion into the Central Cavity}
\label{sec:structure}
\begin{figure}
  \includegraphics[width=\columnwidth]{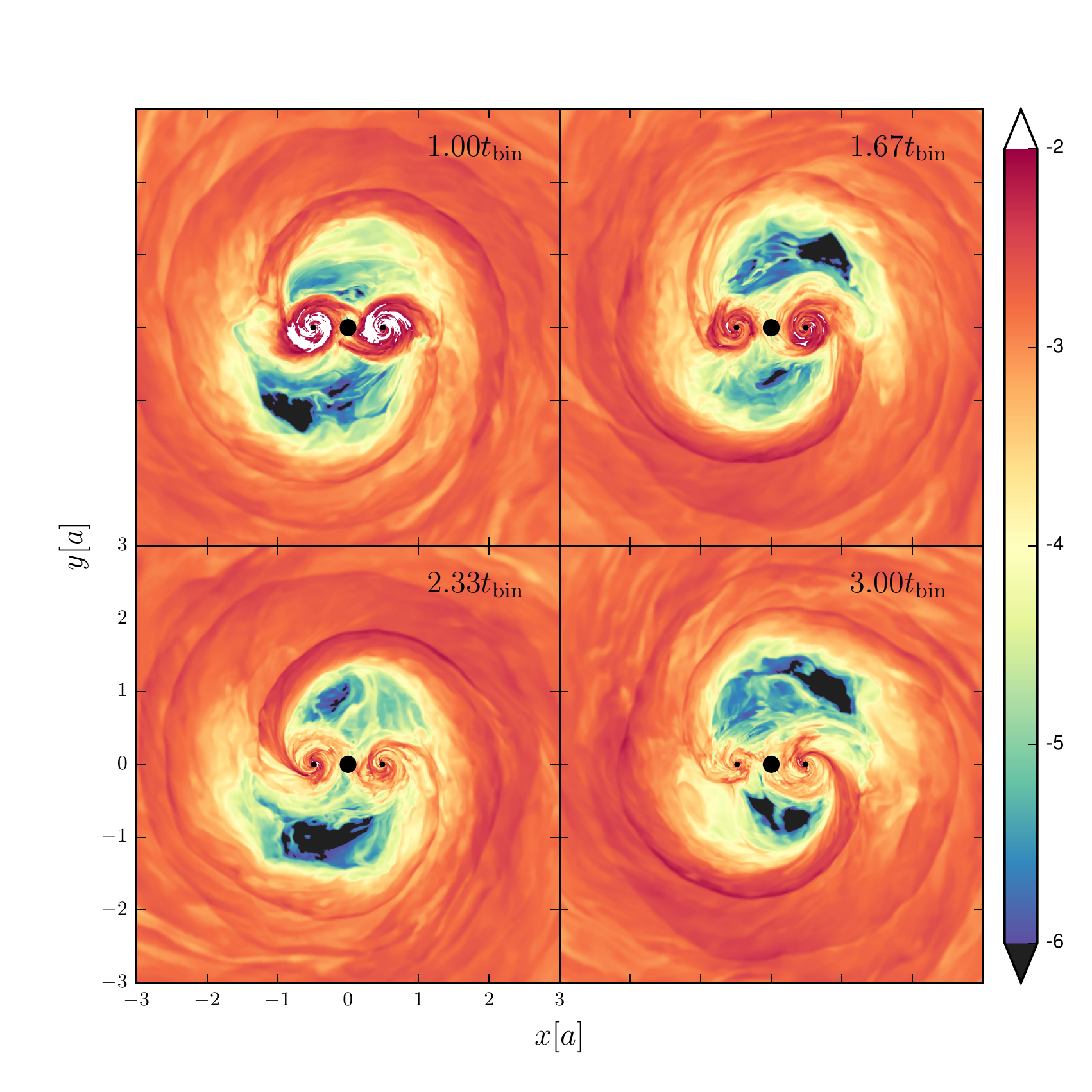}
  \caption{Logarithmic density contours in the equatorial plane of the binary. 
    Simulation time in units of the binary's orbital period is denoted in the top right of each frame. 
    Each BH is represented by a black circle and the larger black circle at the coordinate
    origin marks the central cutout of the simulation.}
  \label{fig:snapshots}
\end{figure}

The circumbinary disk is turbulent, but supports
a lump at $r\approx2.4a$.
The peak surface density of the lump is nearly $4\times$ the mean surface density
of the circumbinary and extends over ${\approx}1$ -- 2~radians in azimuth.
Within $r\approx2a$, there is a low-density cavity across which the streams travel
toward the mini-disks.  These features are shown in Figure~\ref{fig:snapshots}, which
portrays, in a frame corotating with the binary, the gas density in the equatorial plane
at intervals of ${\approx}(2/3)t_{\rm{bin}}$, where $t_{\rm{bin}}=2\pi/\Omega_{\rm{bin}}$.  In this frame, the
accretion streams are almost fixed in position, although their mass content fluctuates.

The times of the four images in Figure~\ref{fig:snapshots} were chosen
so as to highlight a striking effect of the stream asymmetry.  From one to the next,
the BH receiving the most matter changes.  It is BH2 (on the left) in the
first frame, BH1 (on the right) in the next, back to BH2 in the third, and BH1 in the fourth.
This alternation has its origin in the stationarity of the $m=2$ accretion stream pattern
in the corotating frame.  Relative to this stationary pattern, the lump moves backward,
so that it feeds first one stream and then the other.

Previous MHD simulations have shown that the lump orbits at the local
Keplerian frequency,
$\Omega_{\rm{lump}}=\Omega_{K}\left(r=2.4a\right)=0.26\Omega_{\rm{bin}}$~\citep{Shi12,Noble12}.
Therefore, one or the other of the BHs comes into phase with the lump at a frequency
$2\left(\Omega_{\rm{bin}}-\Omega_{\rm{lump}}\right)=1.48\Omega_{\rm{bin}}$. 
Each individual BH does so at precisely half this frequency, $0.74\Omega_{\rm{bin}}$.

\begin{figure}
\includegraphics[width=\columnwidth]{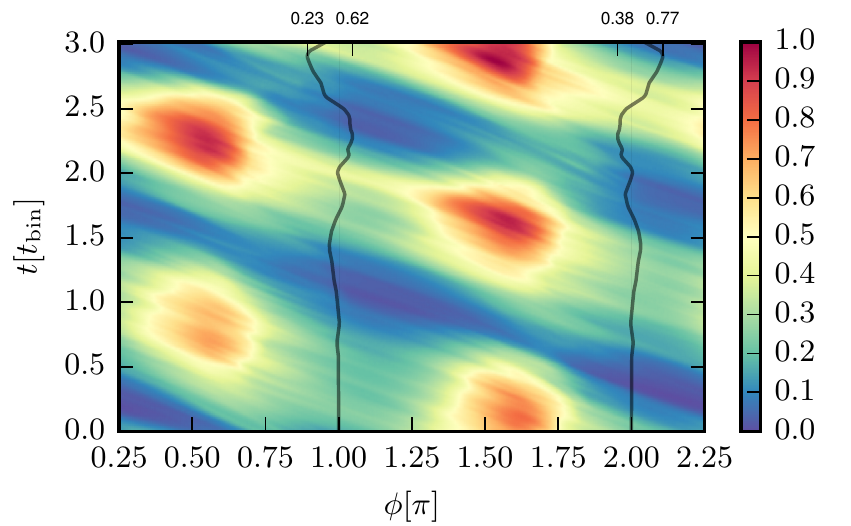}
\caption{(Color Contours) Time and azimuthal dependence of the radially and 
  poloidally-averaged azimuthal mass-flux $\left(\rho v^{\phi}\right)$, normalized
  to the peak value in the simulation, in the frame
  corotating with the binary. $v^{\phi}$ denotes the 
  azimuthal component of the 3-velocity.
  Radial and poloidal averaging is taken for
  $r\in\left(1.5a,2.0a\right)$ and $\theta\in\left(0.4\pi,0.6\pi\right)$
  respectively. Azimuthal angle is shifted
  by $0.25\pi$ so that the stream accreting onto Disk 1 ($\phi=2\pi$)
  and Disk 2 ($\phi=\pi$) are always to the left of the disk for clarity.
  (Vertical Lines) The BH locations are marked by straight lines.
  Semi-transparent
  lines denote the fraction of combined mini-disk mass as a function of time
  within each disk, where shifts to the left or right correspond to less than and 
  more than half the mass respectively. Individual labels at the top provide the minimum and
  maximum value for scale.}
\label{fig:stream_flux}
\end{figure}

We illustrate this switch in preferential accretion more quantitatively in
Figure~\ref{fig:stream_flux} by plotting the radially- and vertically-averaged
azimuthal mass-flux of the accretion streams in a frame corotating with the
binary. This component of the flux provides a good approximation to the mass accretion
rate because in this frame the dominant flow of the streams is in the azimuthal direction.
We find a clear, quasi-periodic oscillation in which the azimuthal location of
the flux switches from near one BH to near the other and back again. The diagonal wedges are
a result of the time delay as the stream moves from the inner edge of the circumbinary to its
target mini-disk. The period of these oscillations is consistent with the
beat-frequency just estimated, $\approx\left(2/3\right)t_{\rm{bin}}$. 

In addition to a quasi-periodic oscillation from one mini-disk to the other,
Figure~\ref{fig:stream_flux} demonstrates that the magnitude 
of the mass-flux in the dominant accretion stream 
increases with time. Conversely, the low flux state between lump-driven
accretion events lack such growth.
Finally, we note that the lump
has been shown to grow secularly over many binary orbits~\citep{Shi12,Noble12}.
Such growth could enhance the observed effect.

\clearpage
\subsection{Depletion and Refilling Cycle of \\Relativistic Mini-Disks}
\label{sec:mini-disk-results-mass}

\begin{figure}
\includegraphics[width=\columnwidth]{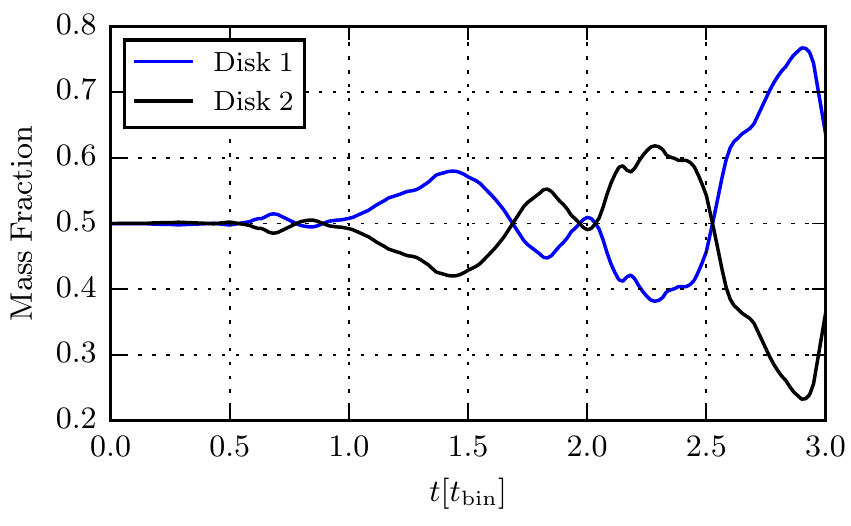}
\caption{Time-dependence of the mass contained within each mini-disk $(r_i\leq0.4a_0)$ normalized
         to the combined mass of both mini-disks.
         Disk 1 and Disk 2 denote the mini-disks around BH1 and BH2 respectively.}
\label{fig:mini-disk-mass}
\end{figure}

The preferential accretion described in Section~\ref{sec:structure} leads to 
two mini-disks whose internal structures differ and never
approach steady-state.  A simple global measure of these facts is the
time-dependence of the mini-disks' masses. In Figure~\ref{fig:mini-disk-mass} we
plot the fraction of total mini-disk mass contained within each disk.
The mini-disks' mass fractions oscillate, while
their contrast continually grows; shortly before the end of the simulation
the mini-disk around BH1 contains over three times the mass of the other mini-disk. 
 
The first two panels of Figure~\ref{fig:snapshots} are at $t/t_{\rm{bin}}=1.0,1.67$,
times when, according to Figure~\ref{fig:mini-disk-mass}, the two disk masses are
nearly equal; the other two panels are at $t/t_{\rm{bin}}=2.33,3.0$, when the
disk masses are near their peak contrasts. 
The more massive stream initially feeds the less massive mini-disk while the other mini-disk drains.
After some time, as seen in the latter two snapshots, the more massive stream's mini-disk becomes the larger one.
In addition, at the later
times, the denser stream extends further into the mini-disk it feeds.  In fact,
after $t\approx2t_{\rm{bin}}$, the stream deposits its material onto the mini-disk
at radii well within the tidal truncation radius, sometimes reaching almost to the
ISCO.
Further evidence for the connection between preferential disk feeding and mass
contrast can be drawn from comparing Figures~\ref{fig:stream_flux} and \ref{fig:mini-disk-mass}.
The times of greatest mini-disk mass contrast coincide with the times of greatest stream
mass-flux contrast.

If the inflow time within a mini-disk were well-predicted by
the classical $\alpha$ model (as assumed in~\cite{Farris14,Farris15a,Farris15}),
it would be
\begin{equation}
t_{\rm{inflow}}\sim[\alpha(H/R)^2\Omega_t]^{-1}\sim80(\alpha/0.1)^{-1}[(H/r)/0.1]^{-2}t_{\rm{bin}},
\end{equation}
where $\Omega_t$ is the local orbital velocity at the outer edge of the mini-disk.
Because the mini-disks in our simulation change in mass on much shorter timescales,
$\sim0.5t_{\rm{bin}}$, it is clear that other mechanisms must be at work.
We discuss what they might be in Section~\ref{sec:discussion}.

\subsection{Spiral Density Waves}
\label{sec:mini-disk-results-spiral}

As Figure~\ref{fig:snapshots} makes plain, strong spiral shocks are a persistent
feature of these mini-disks.  Unlike those seen in the purely hydrodynamic simulations of
\cite{RyanMacFadyen17}, they are highly irregular and non-steady.  This contrast is in keeping
with the Newtonian results of~\cite{Ju16}, who demonstrated that MHD turbulence can have this
effect on spiral shocks in binary systems.  Spiral waves can be excited near Lindblad resonances
in mini-disks by a binary companion's tidal gravity~\citep{Goldreich1979,PapaloizouLin1984,SPL94}.
They are of special interest because once they become nonlinear and steepen into shocks, they can
exert stresses on the fluid~\citep{Papaloizou95,Goodman2001,Heinemann2012,Rafikov16};
where they travel slower than the orbital velocity of the fluid, these stresses transport angular
momentum outward, facilitating accretion~\citep{Ju16,Mewes16b}.

In Newtonian gravity~\citep{SPL94,Makita00,Ju16} and in relativistic
simulations using a Newtonian perturber~\citep{RyanMacFadyen17}, the
spiral patterns have pure $m=2$ character because the dominant
azimuthal Fourier component of the Newtonian perturbing potential is
$m=2$~\citep{SPL94}. However, the perturbing potential in our
relativistic binary spacetime has a significant $m=1$
contribution~\citep{Bowen17}.

To quantify the non-axisymmetric mode strength, we 
compute the normalized azimuthal Fourier amplitudes of the density~\citep{Zurek1986,Heemskerk1992}.
\begin{figure*}
  \includegraphics[width=0.5\textwidth]{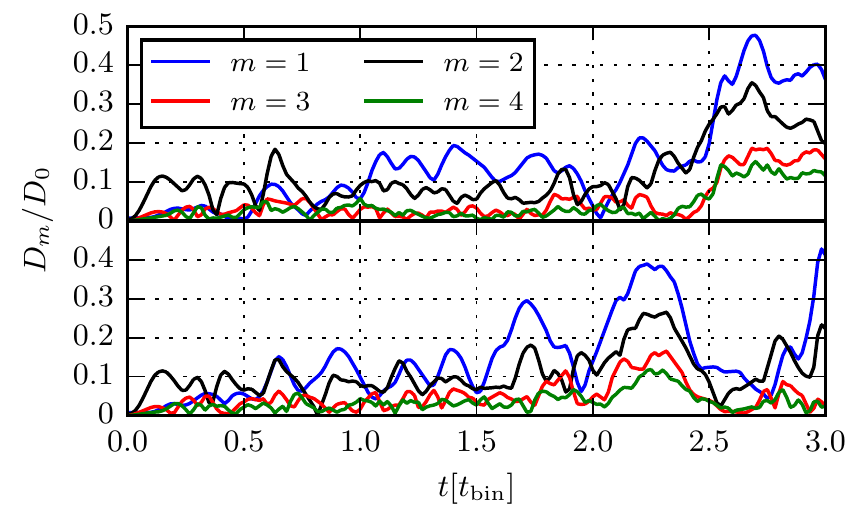}\includegraphics[width=0.5\textwidth]{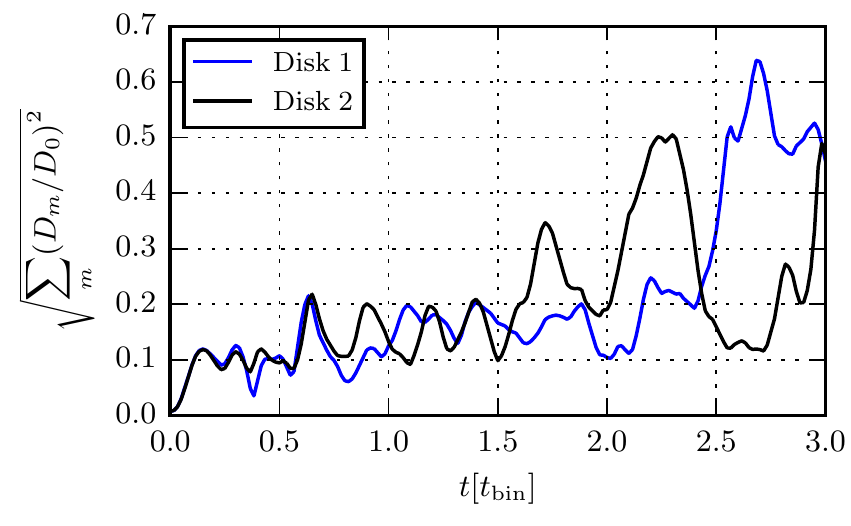}
  \caption{(Left) Time-dependence of the $m=1$ to $m=4$ modes of the density in (Top) Disk 1
  and (Bottom) Disk 2. (Right) Time-dependence of the total mode strength, up to the $m=4$, 
  within each mini-disk. }
  \label{fig:spiral_modes}
\end{figure*}
Figure~\ref{fig:spiral_modes} shows the resulting mode strengths 
for each mini-disk and the total mode strength. 
As found by~\cite{Bowen17}, the dominant Fourier mode is $m=1$. However,
unlike~\cite{Bowen17}, who found significant contributions from only $m=1$ and $m=2$, we
observe lesser, but noticeable, $m=3$ and $m=4$ contributions.  Moreover, the amplitude of all
disk modes grow in time. Interestingly, the total mode strength in each
mini-disk is greatest when its mass is greatest.

In the classical picture, spiral waves are excited only 
by tidal gravity.  However, spiral waves in accreting mini-disks may be incited by pressure waves arising from the impact of accretion streams.  Accretion streams
also alter the spiral wave character by raising the local gas temperature when they strike the
mini-disks: where the ratio of gas temperature to virial temperature is higher, spiral
waves are less tightly-wound and can propagate inward to smaller radii~\citep{SPL94,Ju16}.

\section{Discussion and Conclusions}
\label{sec:discussion}

\subsection{Disk-Mass Oscillations in Relativistic Binaries}

Analytic accounts of gas flow in binaries with near-unity mass-ratios
initially assumed mini-disks would not exist at all~\citep{P91,MP05}; more recent
analytic work assumed that that they would have near-equal masses and maintain a
state of inflow equilibrium~\citep{Roedig:2014}.  Some simulations
\citep{Farris14,Farris15a,Farris15} have found results consistent with these expectations.
In sharp contrast, in our equal mass-ratio simulation, the mini-disks'
masses are in general quite different from one another and are far from inflow
equilibrium at all times.

The reason why our results are so different stems largely from
the relativistically close separation of the binary we treat. The outer edge of
the mini-disk in an equal-mass binary, even in the relativistic regime, is
$r_t\approx0.3a$~\citep{Bowen17}. For a binary separation of $20M$ and a non-spinning
BH, $r_t\approx2.4r_{\rm{ISCO}}$ ($r_{\rm{ISCO}}=2.5M$ in PN harmonic coordinates);
this close to the ISCO, the inflow
rate is no longer well-described by the classic estimate $\sim\alpha(H/R)^2\Omega$.
In fact, explicit simulation of relativistic accretion onto single black
holes has shown that in this regime inflow is considerably faster:
at $r/r_{\rm{ISCO}}\simeq2$,~\cite{KHH05} found an inflow time of only $\simeq7$
local orbital periods for a disk in which $H/R\simeq0.15$ (similar to our mini-disks).
Assuming Keplerian orbits, this inflow time corresponds to
$\approx1.5t_{\rm{bin}}$ for our simulation.  Part of this
inflow acceleration is due to the significant
radial pressure gradient near the ISCO.  Another part is due to the fact that the
local specific angular momentum in this region is not very much larger than that
at the ISCO.  As a result, less stress is required to move inward (this is the
flip-side of the well-known torque reduction factor~\citep{PageThorne74}).
If the stress nonetheless remains large, as it can be when due to MHD turbulence,
inflow can be accelerated.  In a binary whose mini-disks are relatively hot, as our
mini-disks are, spiral shocks can augment angular momentum transport and make
inflow even swifter.

In a simple model of the stream alternation, the accretion rate onto a mini-disk
might be described as a mean rate plus a sinusoidal modulation of fractional amplitude
$\epsilon$ and frequency $\omega_{\rm{mod}}$.  In such a case, the fractional
amplitude of the oscillations in mini-disk mass would be
\begin{equation}
f_{\rm{mass}}=\frac{\epsilon}{[1+(\omega_{\rm{mod}}t_{\rm{inflow}})^2]^{1/2}}.
\end{equation}
In earlier work~\citep{Farris14,Farris15a,Farris15}, $\omega_{\rm{mod}}t_{\rm{inflow}}$ was $\gg1$,
making $f_{\rm{mass}}\ll\epsilon$, because the binary separation was taken to be large enough
that classical inflow rate estimates applied.  In our case, however, $\omega_{\rm{mod}}=0.74\Omega_{\rm{bin}}$
and $t_{\rm{inflow}}=1.5 t_{\rm{bin}}$, leading to $f_{\rm{mass}}\simeq\epsilon/7$.
However, the actual situation departs from this simple model in two ways: the total
accretion rate onto the binary increases over time, and fluctuations in the structure
of the mini-disks can lead to corresponding fluctuations in $t_{\rm{inflow}}$.  In
particular, as already mentioned, the depletion of the mini-disks, combined with the
increased mass-flow in the streams, permits the streams to reach well within the
nominal truncation radius.  Thus, especially in the second half of the simulation,
the matter actually joins the mini-disk at a radius where $t_{\rm{inflow}}$ is
considerably shorter than $\approx1.5t_{\rm{bin}}$, leading to much larger excursions
in mass than predicted by the simple modulation model.

Although our simulations demonstrate that very large amplitude modulation
of the mini-disk masses and accretion rates can happen, it would be premature to claim
that it is general for SMBBHs with $a=20M$.  If the BHs spin rapidly, the ISCO
would move inward, increasing $t_{\rm{inflow}}$.  Both magnetic stress and spiral shocks
are more effective when $H/R$ is relatively large, and mini-disks with lower accretion
rates may be cooler than those simulated here.
  
In addition, the asymmetry in stream mass-fluxes depends on the amplitude
of the lump, but its dependence on system parameters has not yet been
well explored.  It may decrease in magnitude as mass-ratio $q$ drops from $\sim1$ to
$\sim0.1$~\citep{DOrazio13,DOrazio16}. In binaries with smaller $q$,
accretion into the central cavity preferentially targets the secondary. 
This effect would diminish the alternating contrast between the accretion
rates onto the two mini-disks and instead create a quasi-periodic modulation
of the accretion rate onto the secondary's mini-disk. Nonetheless, if
the separation is small enough that the secondary's mini-disk extends to only a few
ISCO radii, the inflow time could still be short enough for the secondary mini-disk's
mass to follow the modulated accretion rate.

A further potential complication has to do with ``decoupling" dynamics, the
events occurring when the inflow time in the {\it circumbinary} disk 
becomes longer than the
orbital evolution time~\citep{MP05}.  For the parameters of this simulation,
\cite{Noble12} showed that decoupling as it is generally envisioned never happens, so that
accretion continues all the way through to merger.  However, it remains possible that for
circumbinary disks with smaller $H/R$, the binary separation at which ``decoupling" occurs may
be considerably larger than the $\sim 20M$ at which it occurred for their (and our) simulation.
In such a case, the binary can evolve to a state in which its orbital frequency is much
greater than the orbital frequency at the inner edge of the circumbinary disk, likely washing
out accretion modulation due to the lump.
Finally, the ratio $\left(r_t/r_{\rm{ISCO}}\right)$ will continue to shrink
as the binary inspirals toward merger. 
This shrinkage may further enhance the depletion/refilling cycle
up until the point at which distinct mini-disks disappear.

\subsection{Further Implications}
 
Just as their structure is quite different from the usual picture
of disks in inflow equilibrium, the radiation produced by mini-disks in the state
we have found should also be quite different from the predictions of such steady-state
models.  Moreover, their intrinsically transient character makes application of
the traditional $\alpha$ model for local dissipation suspect.  Instead, estimates
of the emitted light require direct use of local heating rates due to MHD turbulent
dissipation and shocks.
A first discussion of the electromagnetic emission from 
the mini-disks we simulated is presented in a companion
publication~\citep{RAYTRACE}.

We conclude by mentioning that in our previous treatment of relativistic
mini-disks in 2-d hydrodynamics, we found substantial ``sloshing", gas exchange between
the mini-disks through the L1 region~\citep{Bowen17} (such exchange can be seen in the first frame of Figure~\ref{fig:snapshots}).  
In our present simulation, however, at times later than shown in that panel,
the large central cut-out intercepted most such streams, removing their mass from the system.
In the absence of such a cut-out, we expect that the sloshing will follow the mini-disk
asymmetry, with more material entering the sloshing region from the
more massive mini-disk than the less massive mini-disk.
This mechanism, given a significant mini-disk mass asymmetry, might diminish the mass
disparity between the mini-disks, but the degree to which it does will depend on the detailed mass distribution and internal flow pattern of the more massive mini-disk.
Using the PATCHWORK framework~\citep{PATCHWORK}, which will eliminate the cutout at
the center-of-mass, we intend in the near future to determine what effect sloshing
may have on mini-disk mass variation.
With that additional information, it will also be possible to explore the consequences for photons radiated by the entire mini-disk system, including the sloshing material.



\section*{Acknowledgments}
We thank Mark J. Avara for a careful reading of this manuscript and for
helpful discussions and suggestions. D.~B. would also like to thank
Brennan Ireland for helpful discussions. We would
like to thank the anonymous referee for the careful reading of this manuscript
and for the helpful comments and questions raised.
D.~B., M.~C., V.~M., and M.~Z. received support from NSF grants
AST-1028087, AST-1516150, PHY-1305730, PHY-1707946, OAC-1550436 and
OAC-1516125.  S.~C.~N. was supported by AST-1028087, AST-1515982 and
OAC-1515969, and by an appointment to the NASA Postdoctoral Program at
the Goddard Space Flight Center administrated by USRA through a
contract with NASA.  J.~H.~K. was partially supported by NSF grants
AST-1516299, PHYS-1707826 and OAC-1516247.  V.M. also acknowledges
partial support from AYA2015-66899-C2-1-P. M.~Z. acknowledges support
through the FCT (Portugal) IF programme, IF/00729/2015.

Computational resources were provided by the Blue Waters
sustained-petascale computing NSF projects OAC-0832606, OAC-1238993,
OAC-1516247 and OAC-1515969, OAC-0725070. Blue Waters is a joint
effort of the University of Illinois at Urbana-Champaign and its
National Center for Supercomputing Applications. Additional resources
were provided by XSEDE allocation 
TG-PHY060027N and by the BlueSky Cluster at Rochester Institute of Technology.  
The BlueSky cluster was supported by NSF grants AST-1028087, PHY-0722703 and PHY-1229173.


\end{document}